# Interference effects between ion induced solvent polarisations in Liquid water

Puja Banerjee and Biman Bagchi*

*Solid State and Structural Chemistry Unit, Indian Institute of Science, Bangalore, Karnataka-560012, India*

*Abstract*

*We calculate the effective force between two oppositely charged and similarly charged ions fixed in water as a function of separation distance R. At short separations, R less than 1 nm, the effective force is vastly different from the $1/\varepsilon_s R^2$ dependence advocated by the screened Coulomb force law (SCFL); $\varepsilon_s$ being the static dielectric constant of the medium. We also find an interesting asymmetry in the force between +ve and -ve ions. This breakdown of SCFL is shown to be due to the persistent interference between the polarizations created by the two charges, in a manner similar to the vortex-antivortex pair formation in XY model Hamiltonian. The distance dependence of dielectric constants, $\varepsilon_s(R)$ extracted from our simulation exhibit interesting features and can be used in future modelling.*

## I. Introduction

The role of solvent in the structure and dynamics of electrolyte solutions has been a central research area of physical chemistry, being active for more than a century [1-10]. In particular, interaction between two charges in a polar solvent plays an important role in many chemical and biological reactions. These studies inevitably use Coulomb's law to model electrostatic interactions among the charges. A notable example is found in Marcus' theory of electron transfer. In an aqueous soltion of two ions at a separation R, the solvent dipoles tend to orient themselves in such a way that it can screen the interaction and when two different ions separated by a distance R act simultaneously on the same solvent molecule, this gives rise to an interference whose effects on interaction is hard to quantify.

In order to reduce the enormous complexity in dealing with the many-body systems with free or bound charges, the interaction between two charges immersed in a dipolar liquid is often modelled by a screened Coulomb's force law (SCFL) where the bare coulomb's potential is divided (screened) by the static dielectric constant of the medium, $\varepsilon_s$. *While the bare Coulomb's force law (BCFL) is exact in vacuum, SCFL is only approximate.* According to Coulomb's law (CL), the expression of the bare force between two ions A and B having charges $q_A$ and $q_B$ separated by a distance R is given by

$$F_{BCFL}(R, q_A, q_B) = \frac{q_A q_B}{R^2} \qquad (1)$$

As mentioned above, it is usual, in the condensed phase, to replace this force by the screened form

$$F_{SCFL}(R, q_A, q_B) = \frac{q_A q_B}{\varepsilon_s R^2} \qquad (2)$$

For similar charges, this force is repulsive and for opposite charges this is attractive, at all separations.

A derivation of SCFL (Eq. 2) can be obtained using the polarisation field created by ions in the dipolar liquid[3,11]. In this model, the polarization field, $\mathbf{P}(\mathbf{r})$ depends on the orientations of solvent molecules at a position r. It produces an electrostatic potential $V_A$ at position A, which is given by

$$V_A = -\int_{r>A} \frac{\mathbf{P}(\mathbf{r}) \cdot \mathbf{r}}{|\mathbf{r}|^3} d\mathbf{r} \tag{3}$$

And the free energy of a polarisation field in a dielectric medium is given by [3,1]

$$\Pi = 2\pi \left(1 - \frac{1}{\varepsilon_s}\right)^{-1} \int_{\substack{|\mathbf{r}-\mathbf{r}_A|>a \\ |\mathbf{r}-\mathbf{r}_B|>b}} [\mathbf{P}(\mathbf{r})]^2 d\mathbf{r} \tag{4}$$

If we have two ions, A and B with charges $q_A$ and $q_B$, the free energy of the system is given by

$$\Pi = 2\pi \left(1 - \frac{1}{\varepsilon_s}\right)^{-1} \int_{\substack{|\mathbf{r}-\mathbf{r}_A|>a \\ |\mathbf{r}-\mathbf{r}_B|>b}} [\mathbf{P}(\mathbf{r})]^2 d\mathbf{r} - \int_{\substack{|\mathbf{r}-\mathbf{r}_A|>a \\ |\mathbf{r}-\mathbf{r}_B|>b}} \mathbf{P}(\mathbf{r}) \left[\frac{q_A (\mathbf{r}-\mathbf{r}_A)}{|\mathbf{r}-\mathbf{r}_A|^3} - \frac{q_B (\mathbf{r}-\mathbf{r}_B)}{|\mathbf{r}-\mathbf{r}_B|^3}\right] d\mathbf{r} \tag{5}$$

The equilibrium polarization, $\mathbf{P}(\mathbf{r})$ is obtained by minimizing the free energy expression with the constrain of the potential from the two charges. In the polarisation field, $\mathbf{P}(\mathbf{r})$, the electrostatic potential difference between point A and point B is given by

$$\Delta V = -\int_{\substack{|\mathbf{r}-\mathbf{r}_A|>a \\ |\mathbf{r}-\mathbf{r}_B|>b}} \mathbf{P}(\mathbf{r}) \left[\frac{\mathbf{r}-\mathbf{r}_A}{|\mathbf{r}-\mathbf{r}_A|^3} - \frac{\mathbf{r}-\mathbf{r}_B}{|\mathbf{r}-\mathbf{r}_B|^3}\right] d\mathbf{r} \tag{6}$$

where $r_A$ and $r_B$ denote the positions of A and B respectively.

According to Euler's variational principle, the polarisation function, $\mathbf{P}(\mathbf{r})$ which minimises the free energy $\Pi$ subject to the requirement of equation(6), can be determined by minimizing the functional

$$F = 2\pi \left(1 - \frac{1}{\varepsilon_s}\right)^{-1} \int_{\substack{|\mathbf{r}-\mathbf{r}_A|>a \\ |\mathbf{r}-\mathbf{r}_B|>b}} [\mathbf{P}(\mathbf{r})]^2 d\mathbf{r} - \int_{\substack{|\mathbf{r}-\mathbf{r}_A|>a \\ |\mathbf{r}-\mathbf{r}_B|>b}} \mathbf{P}(\mathbf{r}) \left[\frac{q_A(\mathbf{r}-\mathbf{r}_A)}{|\mathbf{r}-\mathbf{r}_A|^3} - \frac{q_B(\mathbf{r}-\mathbf{r}_B)}{|\mathbf{r}-\mathbf{r}_B|^3}\right] d\mathbf{r}$$
$$+ \alpha \left[-\int_{\substack{|\mathbf{r}-\mathbf{r}_A|>a \\ |\mathbf{r}-\mathbf{r}_B|>b}} \mathbf{P}(\mathbf{r}) \left[\frac{\mathbf{r}-\mathbf{r}_A}{|\mathbf{r}-\mathbf{r}_A|^3} - \frac{\mathbf{r}-\mathbf{r}_B}{|\mathbf{r}-\mathbf{r}_B|^3}\right] d\mathbf{r} - \Delta V\right]$$
(7)

where α is a constant to be determined. By minimizing this functional with respect to the variation of P(r) we obtain the expression of P(r) that can be used to derive an expression for polarisation induced potential energy difference between two charged spheres A and B

$$E_s = \frac{1}{2}\left(1 - \frac{1}{\varepsilon_s}\right)\left[\frac{q_A^2}{a} + \frac{q_B^2}{b} + \frac{2q_A q_B}{R}\right]$$
(8)

where $q_A$ and $q_B$ are the charges on two spheres with radii a and b respectively and R is the distance between centres of two spheres.

Now, the derivative of $E_s$ gives the polarisation induced force (PIF)

$$F_{PIF} = -\left(1 - \frac{1}{\varepsilon_s}\right)\frac{q_A q_B}{R^2}$$
(9)

*This PIF is attractive for two similar charges and repulsive for opposite charges.* By adding this screening force to the force from bare Coulomb's force law ($F_{BCFL}$), we get the screened coulomb's law force with $\varepsilon_s$ in the denominator. We note in passing that the present discussion parallels that in the celebrated Marcus theory of electron transfer [1, 2, 12], where an expression for solvent reorganisation energy ($E_s$) between two charged spheres is derived by using the same approach. Since this derivation is based entirely on continuum framework and no respect is given to the molecular nature of the liquid, its validity is questionable, especially at small-to-intermediate values of separation R.

Long ago Debye pointed out that SCFL can breakdown close to an ion because the dielectric constant can become position dependent and should be replaced by $\varepsilon_s(r)$ [3], as often practiced in theories and simulations, particularly for biomolecules. The most simple approach is to use $\varepsilon_s(r)$ as a linear function of distance from the ion/charge(r)[13 14], that is, $\varepsilon_s(r) = Cr$, where C is a constant generally assigned with the values ranging from 1 to 4.5. In an influential study using continuum model, Warshel *et al.* developed a more accurate functional form of $\varepsilon_s(r)$ that varies exponentially with distance[15]. Despite a large number of theoretical studies focused on the construction of model dielectric function in different biological systems (Olson and co-workers[16], Hingerty and co-workers[17], Ramstein and Lavery[18]), the rational for such replacement is often not clear, but nevertheless it serves practical purpose in simulations. However, all these continuum model descriptions neglect the molecular nature of the solvent.

There is a different, not quite anticipated, consequence of this neglect of molecularity. Due to the long range nature of polar interaction, the polarisation induced by any of the changes may persist up to a long distance. In a fascinating experiment, Roke *et al.*[19] recently demonstrated the existence of such correlation between water at a large length scale (even beyond 8 nm) in presence of ions. To explain this experimental study, Laage, Jungwirth and coworkers[20] carried out an interesting simulation study that seems to show that the long-range order of water-water orientational correlation function could indeed be influenced by ion-ion correlation, as it affects their individual solvation shells. The interference of polarization created by two ions that modify the force between them was not investigated. An example of such intererence was discussed in Ref. 21 that considered water within a reverse micelles. The interference can be destructive (for two positive ions) and constructive (for oppositely charged ions).

Earlier in a mode-coupling theory analysis of diffusion and viscosity, it was pointed out that the molecular nature of ion-solvent interaction can play an important role even at low concentrations[22-24]. It was also suggested that local orientational correlations can be used to obtain a wavenumber dependent dielectric function, $\varepsilon(k)$, using classical density functional theory [25].

In the present work we extract distance dependent dielectric function, $\varepsilon_s(R)$ from a full atomistic simulation of an ion pair in water where R is the separation distance of the ion-pair.

Our simulations reveal that *screened Coulomb's potential loses its validity at distances less than 1-1.5 nm in liquid water, for monatomic ions* The force between two similar charges is found to be more repulsive and that of two opposite charges more attractive by a large factor even in the absence of any other ionic species in the system. We also find an asymmetry in the polarization distribution around an anion and a cation which could have detectable effects.

## II. System and simulation details:

Molecular dynamics simulations of potassium chloride (KCl) in water have been carried out using the Lammps package [26]. Rigid non-polarizable force field parameters have been used for water as well as ions. SPC/E model[27] has been employed for water. For ions, potential parameters from Ref. [28] have been employed. The self interaction parameters are listed in Table 1 and consist of Lennard-Jones and Coulombic terms.

We have taken two types of systems. In the first type, we have taken 1 $K^+$ and 1 $Cl^-$ in a cubic simulation cell of length 70.0 Å with 11319 water molecules and the positions of $K^+$ and $Cl^-$ have been kept fixed with different distances between them varying from 3 to 15 Å.

In the second type, we have taken two K$^+$ ions in the same simulation cell and the positions of two ions have been kept fixed with different distances between them. This is a system with no negative charges to see only the effect of water on the interaction of two K$^+$ ions.

Constrained MD simulations were carried out by fixing the positions of ions at a particular distance of separation in the microcanonical ensemble with periodic boundary conditions with a cut-off radius of 15 Å. The long-range forces were computed with Ewald summation[29,30]. Trajectory was propagated using a velocity Verlet integrator with a time step of 1 fs. The aqueous KCl system was equilibrated for 300 ps at 298 K and then a 6 ns MD trajectory was generated in the microcannonical (NVE) ensemble. The coordinates were stored every 10 fs for subsequent use for the evaluation of various properties. Finally, we have reported the results averaged from several different trajectories.

**Table 1: Values of Lennard-Jones and electrostatic interaction potential parameters. e represents the magnitude of the electronic charge.**

| Atom, i | $\sigma_{ii}$ (Å) | $\varepsilon_{ii}$ (kJ/mol) | $q_i$ (e) | Ref |
|---|---|---|---|---|
| H$^w$ | 0.000 | 0.000 | +0.4238 | 27 |
| O$^w$ | 3.169 | 0.6502 | -0.8476 | 27 |
| K$^+$ | 3.331 | 0.4184 | +1.0 | 28 |
| Cl$^-$ | 4.40 | 0.4184 | -1.0 | 28 |

# III. Results and Discussion

## a. Polarisation field created in water around an ion

It is evident that the water molecules located near a given ionic species become oriented in a particular direction due to ion-dipole interaction. Polar solvents, other than water, such as methanol, DMSO etc. also exhibit such rearrangements of solvent dipoles around charged species. We have defined a molecular dipole of water molecule and measured the angle between the dipole vector and the ion-water connecting vector ($\theta$) (**Figure 1(a)**). Due to the nature of ion-water interaction, polarisation measured by $\cos\theta$ possesses inverse profiles for the two systems of similarly charged ions and oppositely charged ions but is not actually a mirror image of each other. This has already been suggested in earlier studies that the water solvation shell around a positively charged ion is more structured than that around a negatively charged ions. The polarisation profiles exhibit two major peaks and having higher polarity, polarisation of water extends slightly larger than 10 Å ( see **Figure 2(b))**.

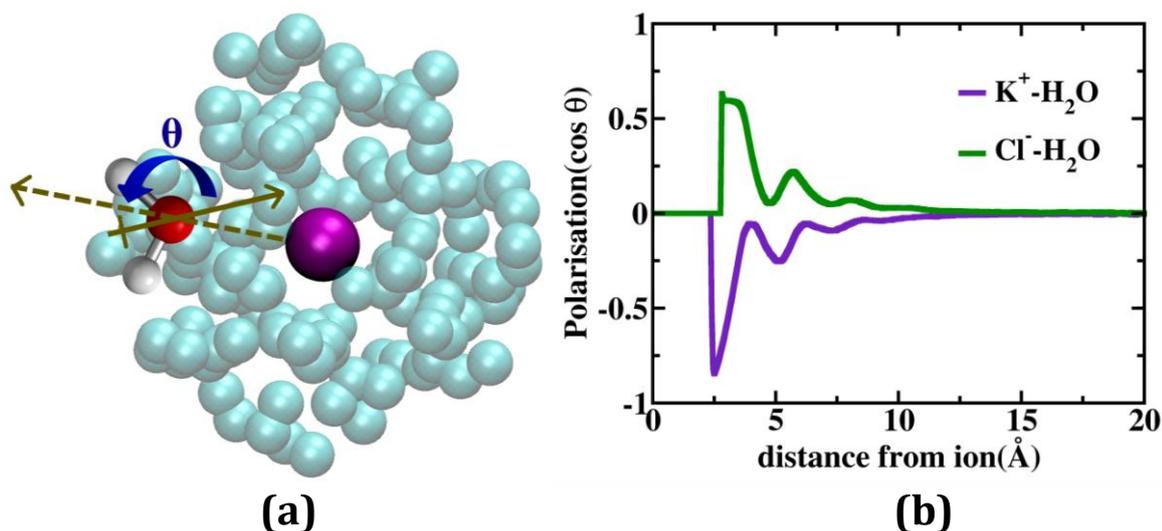

(a)            (b)

**Figure 1: (a) Hydration shell around an ion and representation of the polarization angle between water dipole and ion-water intermolecular axis, (b) Polarisation profiles of water around isolated positive and negatively charged ions.**

## b. Interference between the two induced polarisations in water

We have carried out a series of long atomistic simulations with ion pairs fixed in water with constrained molecular dynamics at different distances of separation between them. Our main aim is to analyse the polarisation of water in the presence of two fixed charges and the modified force between two charges due to the polarisation force. The resultant value of polarisation along the inter-ionic axis is determined by the constructive interference of polarisation (**2(a))**. Due to this, the polarisation of water between two oppositely charged ion with a shorter inter-ionic distance increases, that it turn modifies the force acting between two oppositely charged ions that we shall discuss later in detail. In **2(b)** the average dipole vectors of water molecules is plotted around two ions, $K^+$ and $Cl^-$. *The situation is tantalizingly analogous to the formation of vortex-antivortex pair as found in XY model.* It is evident that the water molecules located in between the two ions (near to the inter-ionic axis) are oriented in the same direction through out the inter-ionic region.**Figure 2(c)** shows the polarisation profiles in the presence of two oppositely charged ions obtained from the simulation. Here polarisation has a structural pattern, hence, the unique nature of constructive interference at different distance of separation finally dictates the attractive force between them.

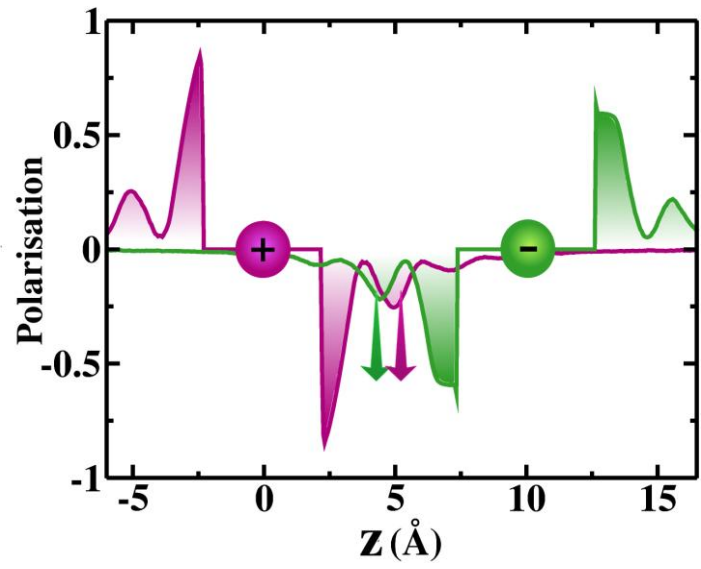

(a)

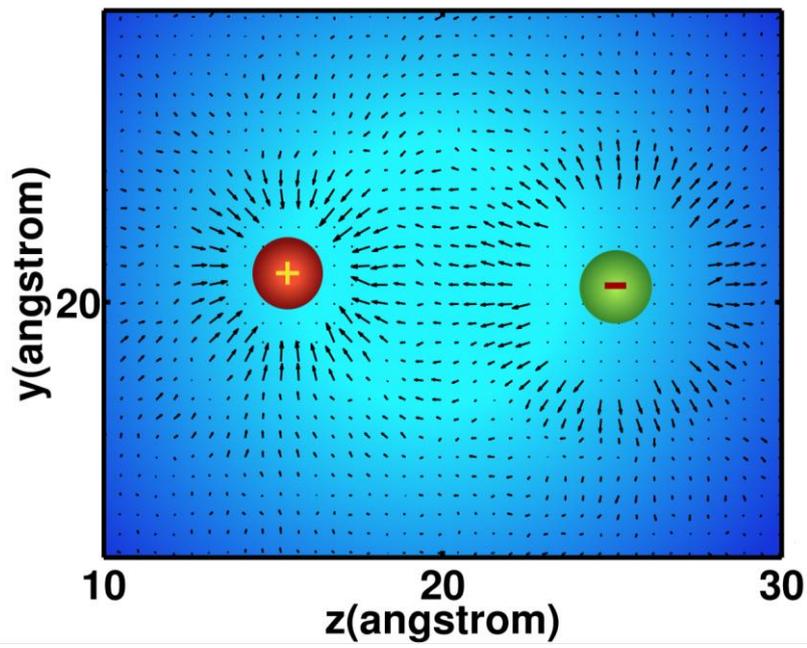

(b)

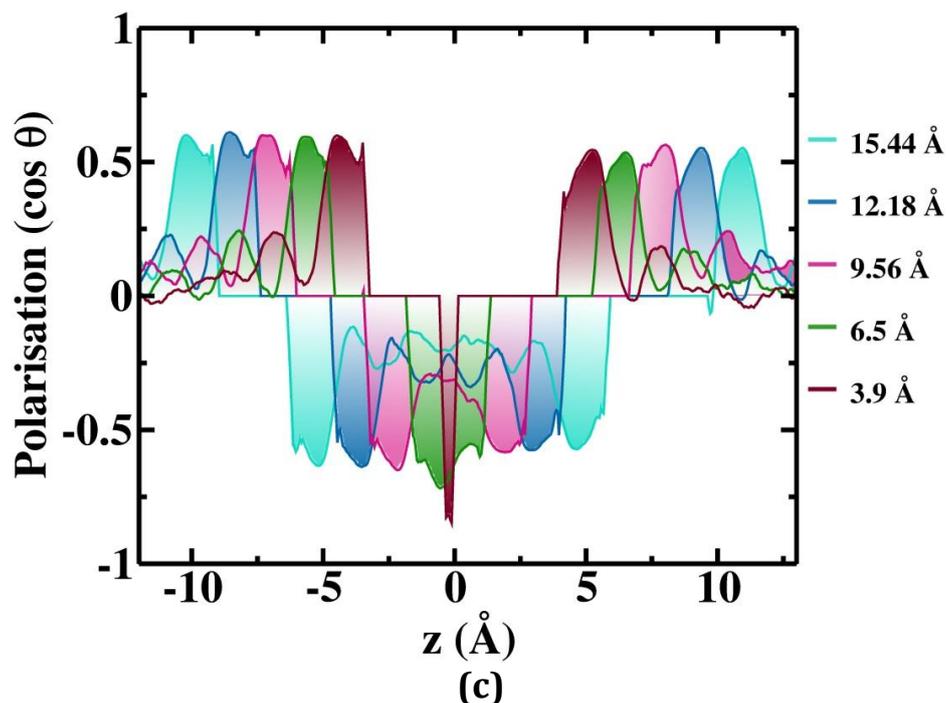

**Figure 2:** (a) The polarisation interaction of water in the presence of two oppositely charged ions at a separation of 10 Å, (b) Average direction of dipole vectors around that $K^+$ and $Cl^-$ ion-pair. (c) Polarization profiles obtained from simulation along the inter-ionic axis of ion-pair.

Next, we focus on two positively charged ion-pair. Here, polarisation of water molecules between two ions along the inter-ionic axis experience a destructive interference(**Figure 3(a)**). **Figure 3(b)** shows that the dipole vectors in the central region of inter-ionic axis does not have any particular reorientation due to a lower polarisation of water molecules. This modifies the force between two positively ions such that at shorter separation they experience much more repulsion than that predicted by screened Coulomb's law. **Figure 3(c)** shows the polarisation profiles in the presence of two positively charged ions obtained from the simulation.

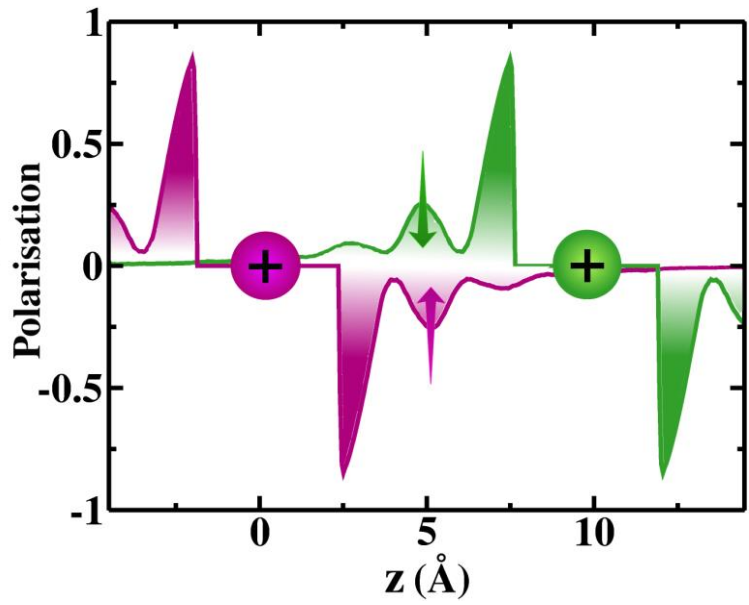

(a)

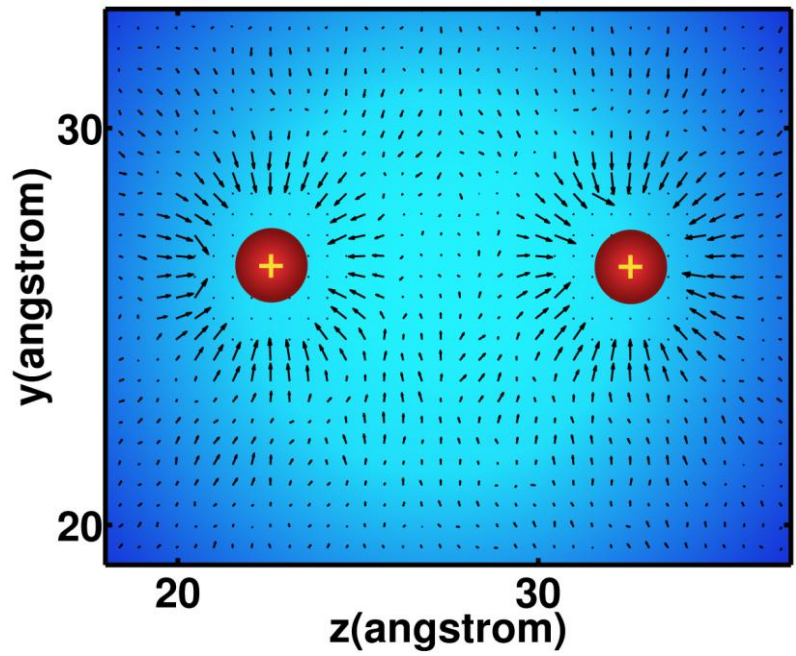

(b)

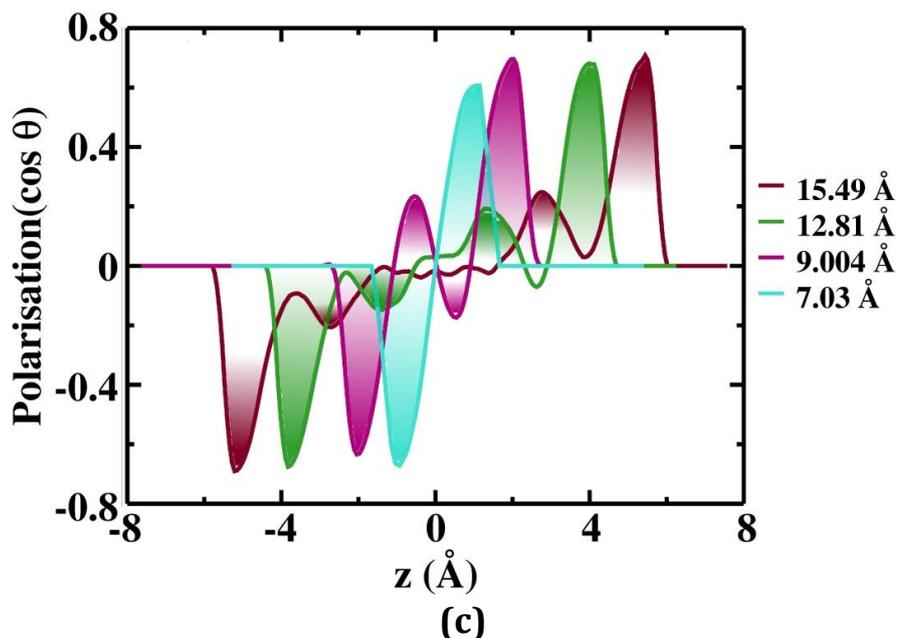

**Figure 3:** (a) The polarisation interaction of water in the presence of two positively charged ions at a separation of 10 Å, (b) Average direction of dipole vectors around the two $K^+$ ions, (c) Polarization profiles obtained from simulation along the inter-ionic axis of the ion-pair.

## c. Effective force between two charges in water

We have computed the polarisation-induced force acting between the two ions and add this to the bare Coulomb's force the effective total force between two ions is obtained. Towards this, we first measure the total electrostatic force on $ion_1$ exerted by all the water molecules ($F_1$) and the total force on $ion_2$ exerted by all water molecules ($F_2$). Then we project $F_1$ and $F_2$ on the inter-ionic axis and get the projected forces $F_{1P}$ and $F_{2P}$ respectively. Finally, we have taken the average of $F_{1P}$ and $F_{2P}$ as the polarization induced force ($F_{PIF}$) between the two ions which is repulsive for two oppositely charged ions and attractive for two similar ions, therefore screens the Coulomb's force between them. The effective force ($F_{eff}$) between two ions can be obtained by adding this force with the Coulomb's bare force ($F_{BCFL}$). As $F_{PIF}$ and $F_{BCFL}$ are similar in magnitude (for water) but opposite in sign, the effective force is smaller by a factor of ~50-100 which makes it difficult to get the values accurately. To overcome this, we created several long trajectories with frequently saved data to generate sufficient data points to obtain accurate average. A pair of $K^+$ and $Cl^-$, with positions fixed in water and the distance of separation is changed within a range 3-15 Å in a series of simulations and the different contributions to the effective force between them is shown in **Figure 4(a).** Here the polarization force is found to be repulsive that screens the attractive Coulomb's force between $K^+$ and $Cl^-$. **Figure 4(b)** shows the effective distance-dependent force for the system of two $K^+$ ions in water where the polarization induced force, $F_{PIF}$ is seen to be attractive here that screens the Coulomb's repulsive force between two positive ions and gives the effective force between them.

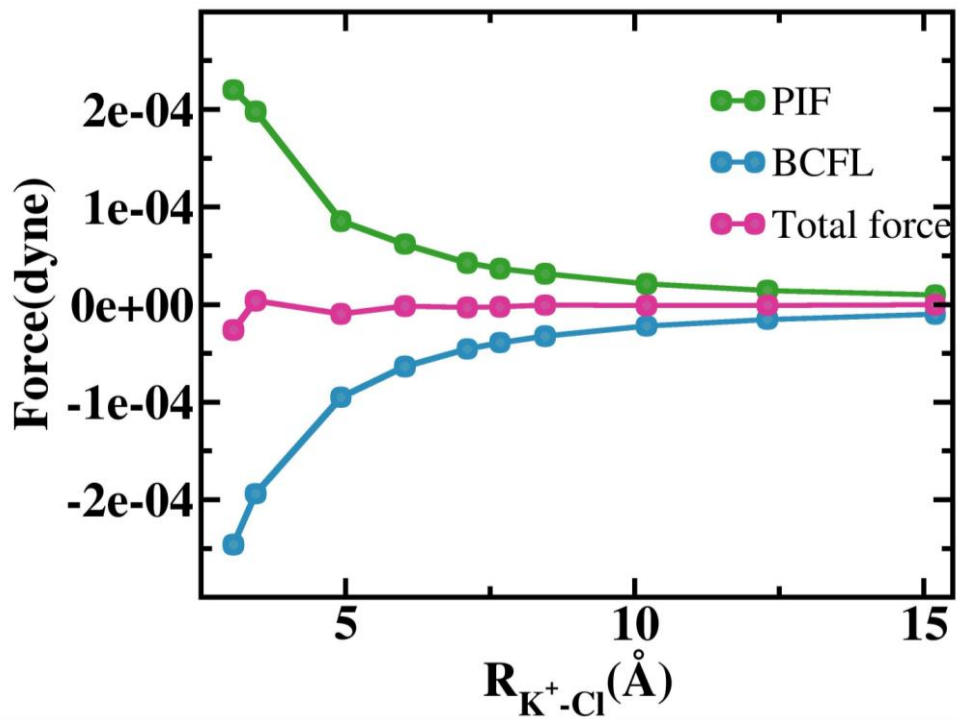

(a)

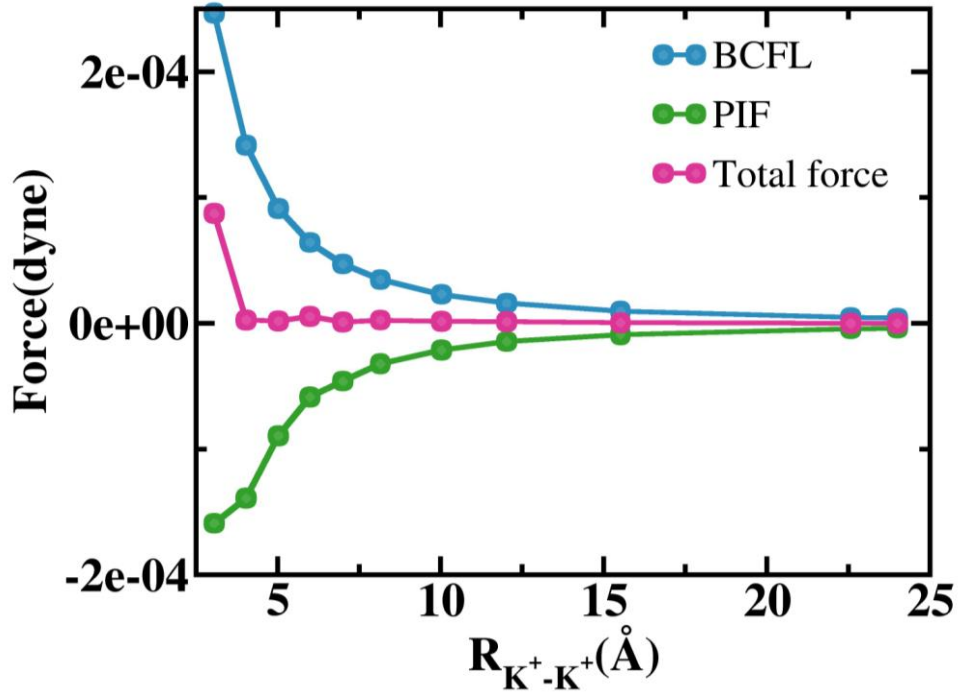

**(b)**

**Figure 4:** (a) Distance dependence of forces between K$^+$ and Cl$^-$ ions in water from bare coulomb's force law(BCFL), the polarization induced force(PIF) from simulation and the total force between them, (b) corresponds to the system of two K$^+$ ions in water.

The total effective force from simulation is found to have oscillatory non-monotonic nature and it is more attractive for certain range of separation distance typically below 10 Å. The distance-dependence of the screened force ($F \propto \dfrac{1}{R^2}$) is modified. Interestingly, the distance dependence of dielectric function, $\varepsilon_s(R)$ extracted from the effective force **(Figure 5(b))** exhibit a wave-like variation with a magnitude substantially lower than 60 at separation distance less than 10 Å. Similarly, the effective force between two positive charges is more repulsive than the SCFL(**Figure 5(c)**) and the extracted $\varepsilon_s(R)$ remains much reduced upto 15 Å separation of two K$^+$ ions (**Figure 5(d)**).

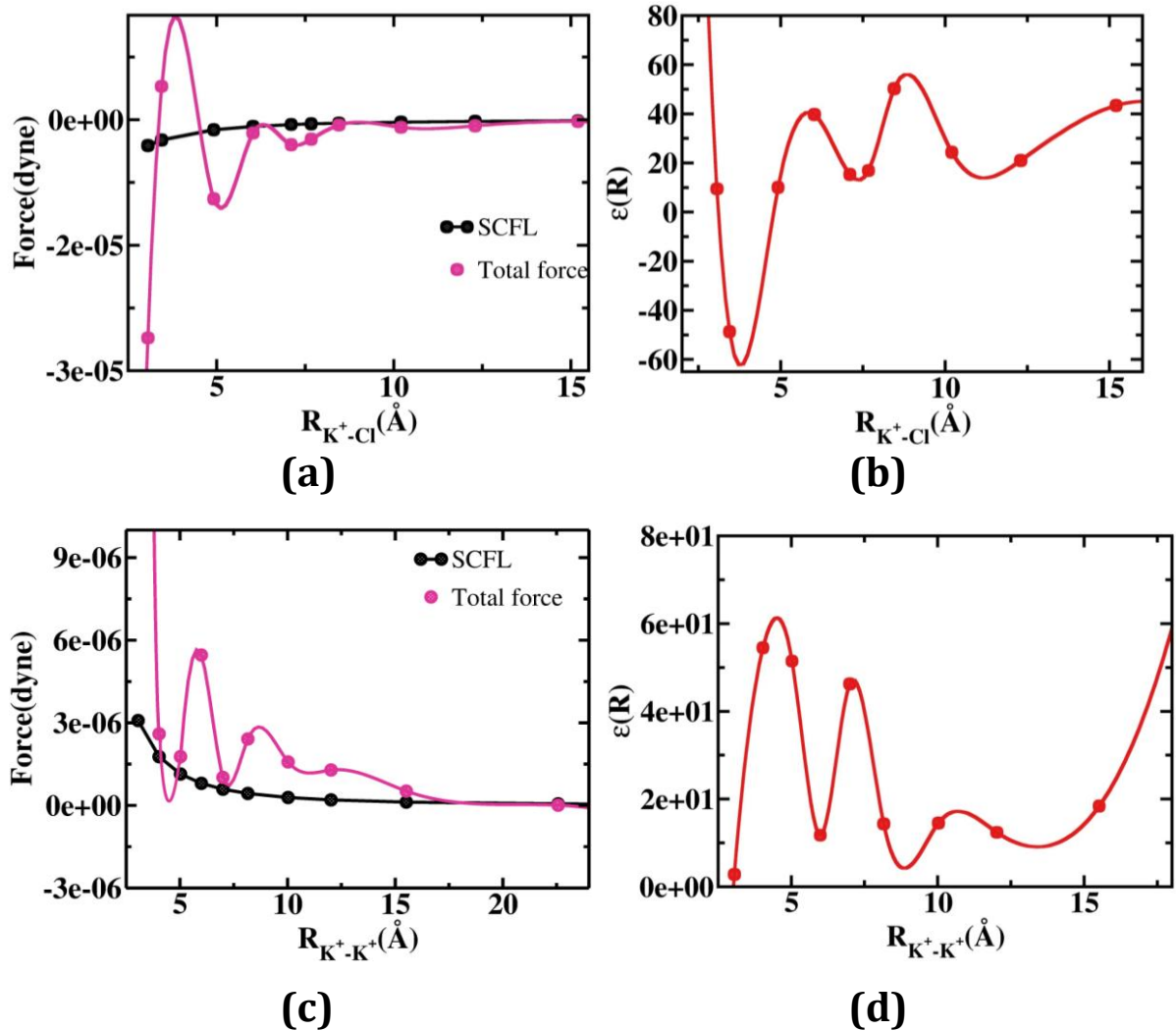

Figure 5 : (a) Comparison of the effective force between two ions (K$^+$ and Cl$^-$) calculated from simulation with that predicted by the screened Coulomb's force law(SCFL), (b) Distance dependent dielectric function, $\varepsilon(R)$ between the ion-pair. *Please note that usually r is used to define distance dependent dielectric constant everywhere in literature. We have deliberately changed the notation as here distance between two ions is denoted by R*, (c-d) correspond to the system of two K$^+$ ions in water.

## d. Potential Mean Force(PMF) between two ions in water

We have computed potential of mean force(w(r)) between two ions in water that is defined as

$$w(r) = -k_B T \ln g(r) \qquad (10)$$

We have used radial distribution function between three ion pairs (K$^+$-K$^+$ and K$^+$-Cl$^-$) in a system of 1M aqueous KCl salt solution and obtained the PMF between them. The effective force here includes the contributions from other ionic species and modified polarisation force due to their presence. This distance-dependent interaction potential is a mean-field result which differes significantly from the effective force(s) between two charges we have calculated from our simulation.

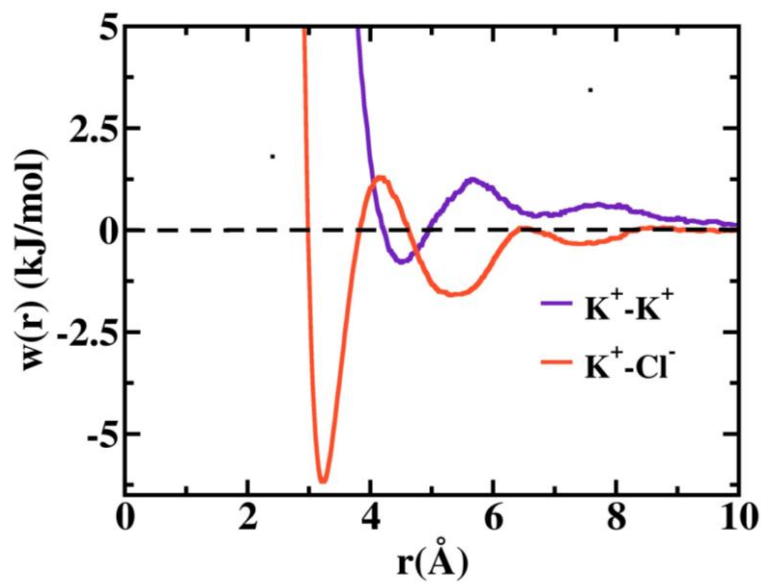

(a)

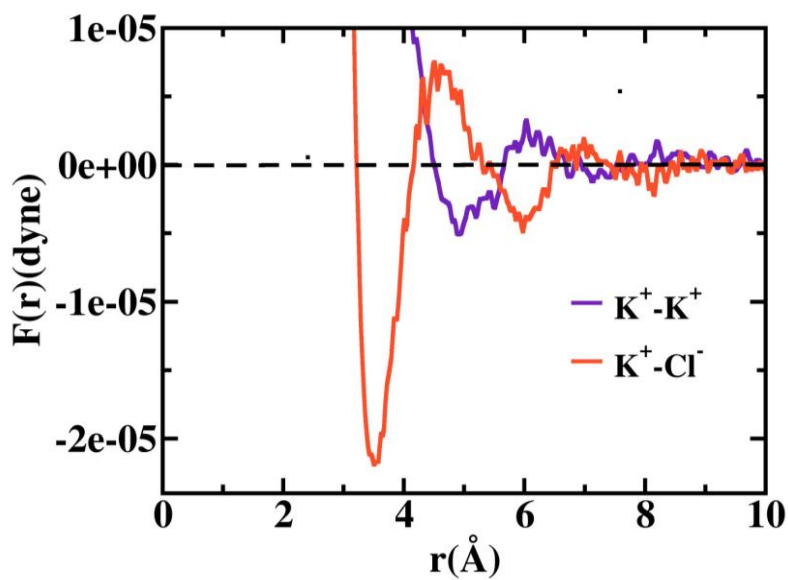

(b)

**Figure 6: (a) Potential mean force (PMF) for different ion-pairs obtained from the simulation of 1M aqueous KCl salt solution, (b) Effective force between ion-pairs derived from PMF.**

Figure 6 and Figures 4-5 deserve a comparative discussion. Although there are certain degrees of similarity between them, the two are not the same. There are differences in positions of minima and maxima, also in depths and heights and the lengthscale. Note that we should not expect a close agreement because the potential of mean force captures an approximate description including the effects from all the ions present in the system.

## E. Liquid Density and mean-square displacement of water between two ions

We have computed the density profile of water molecules between two oppositely charged ions. At a separation of ~5 Å, there is only one layer of water molecules though the number density is greater than the bulk density (shown by the dotted lines in **Figure 7(a)**). With the increase of distance of separation, the profile also changes and after ~10 Å the density profile in the central region of interionic axis becomes similar to the bulk density. The asymmetry found in the two peaks near the ions is real. The density profile reflects formation of structures in the water between the two ions.

We have also analysed the dynamics of those water molecules between $K^+$ and $Cl^-$ ions. As the polarization of water molecules between two oppositely charged ions increases and the water molecules become more structured, the diffusivity of water molecules also decreases significantly (**Figure 7(b)**). These differences are expected to play a role in determining the rates of various chemical processes.

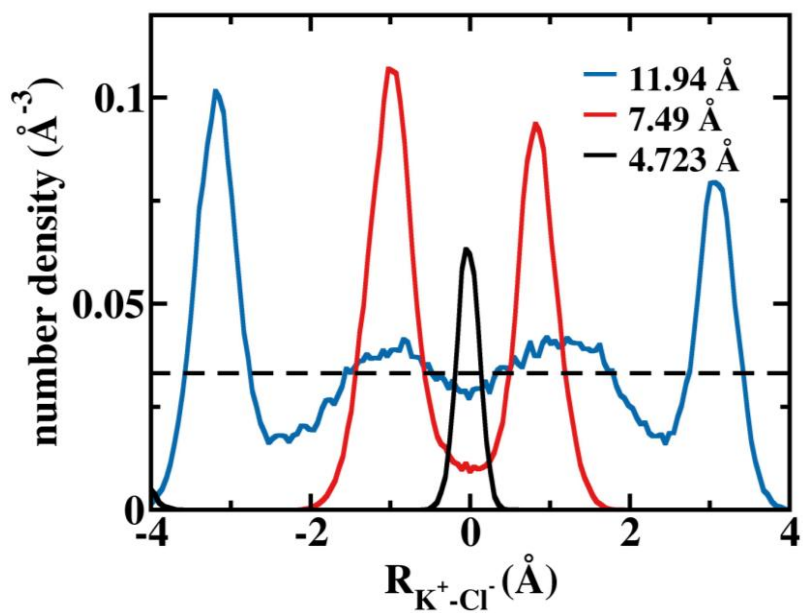

(a)

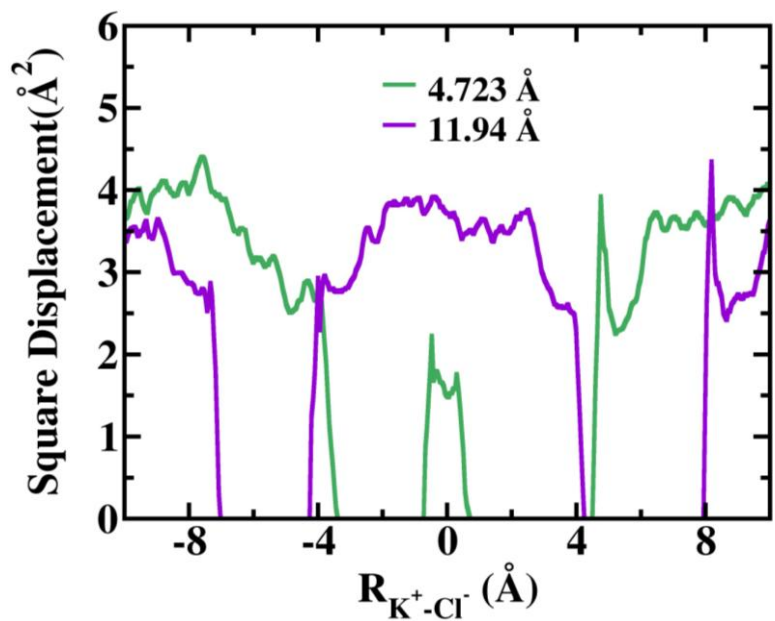

(b)

**Figure 7:** (a) Density profile of water molecules between $K^+$ and $Cl^-$ ion-pair. The bulk number density of water is shown by dotted lines, (b) Square displacement of water around ion-pair for the time-interval 2 ps.

## IV. Conclusions

The dielectric screening effect embodied in Poisson-Boltzmann (or generalized Born) description is used widely in simulations of electrolye solutions and biomolecules. Such treatments assume that each ion experiences an average influence of its surrounding ionic environment. In this description, no ion-ion correlation or fluctuation effect is included that could play an important role in high ion charge density. Similarly, in many theoretical approaches including Debye-Huckel theory, primitive model of electrolyte solutions, mean spherical approximation models etc. also invoke dielectric screening of inter ionic interactions. Therefore, the ability of SCFL to predict the interaction between charges decides the success of such theoretical or computational techniques.

However, any microscopic description must reduce to the screened Coulomb law at large separations. A microscopic expression for polarisation density, P(r) in a dipolar liquid is given by

$$P(r) = \mu \int d\omega \, \hat{\alpha}(\omega) \rho(\mathbf{r},\omega) \tag{11}$$

where $\mu$ is the magnitude of dipole moment of a solvent molecule, $\rho(\mathbf{r},\omega)$ is the position and orientation dependent solvent density and $\hat{\alpha}(\omega)$ is a unit vector with orientation $\omega$. As discussed elsewhere [31], one can use classical density functional theory (DFT) to obtain a free energy in terms of the polarisation field [25, 31]

$$\beta \mathcal{F}[\mathbf{P}(\mathbf{k})] \cong \frac{1}{3y(2\pi)^2} \left\{ \int d\mathbf{k} \, P^2(\mathbf{k}) - \frac{\rho_0}{3} \int d\mathbf{k} \, \mathbf{P}(\mathbf{k}).\mathbf{C}(\mathbf{k}).\mathbf{P}(-\mathbf{k}) \right\} - (2\pi)^{-3} \int d\mathbf{k} \, \mathbf{P}(\mathbf{k}).\mathbf{E}_0(\mathbf{k}) \tag{12}$$

where $E_0(k)$ is the bare electric field of the charged particle, $C(k)$ is related to the wave number dependent static dielectric tensor of the liquid and $3y = 4\pi\beta\rho_0\mu^2/3$. [5] This is the microscopic counterpart of Eq. (5) and reduce to Eq. (8) under appropriate conditions. In this case the free energy functional needs to be minimized with respect to **P(k)** in the presence of the field due to two ions separated by a distance R. This feature makes the evolution of **P(k)** (or, **P(r)**) highly non-trivial. We hope to return to this problem in future.

The present computer simulation based study reveals several interesting aspects of the effective force between two ions in water. We observe an interesting anti-vertex pair like formation of water dipoles around the two +ve ions (shown in Figure 3). The effective force between two ions differs largely from SCFL for the separation distance less than 10-15 Å. The attractive interaction between two $K^+$ ions as seen from PMF is absent in Figure 6: (a) Potential mean force (PMF) for different ion-pairs obtained from the simulation of 1M aqueous KCl salt solution, (b) Effective force between ion-pairs derived from PMF. which

In summary, we have shown that the interference between the polarizations of water owing to the presence of two ions decides the polarization induced force that adds to the bare Coulomb's law and determines the effective force. The resultant force is found to be vastly different from the screened Coulomb's force law at a separation lower than 10-15 Å. Therefore, this study partly explains the failure of the theories like DHO etc. at high concentration and questions the simulation of biomolecules using Poisson-Boltzmann or generalised Born equation etc. for systems with higher charge density. The computed distance dependent dielectric froction, $\varepsilon_s(R)$ reveals that the screening could significantly reduce at intermediate separations, with some multiple minima which leads to the formation of contact ion-pair, solvent separated ion-pair etc. The present result of $\varepsilon_s(r)$ can be used in many simulations.

# Acknowledgement

We thank Department of Science and Technology (DST, India), Council of Scientific and Industrial Research (CSIR, India) and Sir J. C. Bose Fellowship to Prof. B. Bagchi for providing partial financial support